\documentclass[prl,twocolumn,groupedaddress,nofootinbib, superscriptaddress]{revtex4-1}
\pdfoutput=1

\usepackage{amsmath,amssymb,amscd}
\usepackage{listings}
\usepackage{dsfont}
\usepackage{slashed}
\usepackage{color}

\usepackage{hyperref}
\hypersetup{
    colorlinks=true,
    citecolor=blue,
    linkcolor=blue,
    urlcolor=magenta,
}
\usepackage[capitalise]{cleveref}

\usepackage[pdftex]{graphicx}
\usepackage{epstopdf}
\usepackage{subfigure}
\usepackage{epsfig}
\begin{document}

\rightline{}
\rightline{{\tiny Cavendish-HEP-19/08, DAMTP-2019-21, KCL-PH-TH-2019-45}}

\title{Light Dark Matter from Inelastic Cosmic Ray Collisions}


\author{James~Alvey} 
\author{Miguel~D.~Campos}
\author{Malcolm~Fairbairn}

 \affiliation{Theoretical Particle Physics and Cosmology Group, Physics Department, \\
King's College London, London WC2R 2LS, UK}

\author{Tevong~You}

\affiliation{DAMTP, University of Cambridge, Wilberforce Road, Cambridge, CB3 0WA, UK; \\
Cavendish Laboratory, University of Cambridge, J.J. Thomson Avenue, Cambridge, CB3 0HE, UK}

\date{May 2018}

\begin{abstract}
Direct detection experiments relying on nuclear recoil signatures lose sensitivity to sub-GeV dark matter for typical galactic velocities. This sensitivity is recovered if there exists another source of flux with higher momenta. Such an energetic flux of light dark matter could originate from the decay of mesons produced in inelastic cosmic ray collisions. We compute this novel production mechanism---a cosmic beam dump experiment---and estimate the resulting limits from XENON1T and LZ. We find that the dark matter flux from inelastic cosmic rays colliding with atmospheric nuclei can dominate over the flux from elastic collisions with relic dark matter. The limits that we obtain for hadrophilic scalar mediator models are competitive with those from MiniBoone for light MeV-scale mediator masses.
\end{abstract}

\maketitle 


{\it Introduction.} --- While we have a plethora of indirect observational evidence for the existence of some form of dark matter in the Universe, the experimental search for a direct detection signature~\cite{Goodman:1984dc} is still ongoing. Its scope has expanded beyond the weak-scale thermal relic paradigm, and now encompasses a wider variety of possibilities for new physics beyond the Standard Model~\cite{Battaglieri:2017aum}. Light sub-GeV dark matter, in particular, has become a prime target of such activities. 

Acquiring sensitivity to sub-GeV dark matter typically requires detectors able to pick up lower recoil energies; a reduced momentum is expected for lighter masses given that the galactic velocity of dark matter is $\mathcal{O}(10^{-3}) c$. However, Refs.~\cite{Bringmann:2018cvk, Ema:2018bih} recently showed that light dark matter interacting with nucleons or electrons necessarily leads to an energetic flux due to cosmic rays colliding elastically with dark matter in the interstellar medium. This up-scattered dark matter flux may then have enough energy to be detectable in direct detection experiments such as XENON1T (previously thought to be sensitive only to $\mathcal{O}(10-100)$ GeV dark matter)~\cite{Aprile:2018dbl}, as well as other dark matter or neutrino detectors~\cite{Bringmann:2018cvk, Ema:2018bih}~\footnote{See also Refs.~\cite{Cappiello:2018hsu, Kouvaris:2016afs, Ibe:2017yqa, Dolan:2017xbu} for ways of extending the direct detection sensitivity to lighter dark matter masses (in particular Ref.~\cite{Cappiello:2018hsu} for another way of using of cosmic rays) and Refs.~\cite{Kouvaris:2015nsa, An:2017ojc, Emken:2017hnp} for solar sources of energetic dark matter flux.}. 

Here, we point out another generic source of light dark matter flux. If mesons decay partially into dark matter, as could happen through the same coupling to nucleons that enables direct detection, then the mesons generated in inelastic cosmic ray collisions will also produce an energetic flux of dark matter. This may be viewed as a continuous cosmic beam dump experiment. It naturally provides a preexisting light dark matter source for experiments that would otherwise be insensitive to them. The different detector targets, exposure, and source geometry involved then enable distinctive opportunities relative to dedicated beam dump experiments. Indeed, we shall see that XENON1T~\cite{Aprile:2018dbl} and the future LZ experiment~\cite{Akerib:2018lyp} set competitive limits for light mediators in comparison to MiniBoone~\cite{Aguilar-Arevalo:2018wea} (XENONnT, while under construction, has not published projected limits/experimental sensitivities as of yet and therefore is not considered in our study). Our source requires meson decay into light dark matter and is therefore not irreducible like the upscattering mechanism, which only assumes the nucleon interaction cross-section required for detection. However, the upscattering mechanism relies on a relic dark matter density, whereas inelastic cosmic ray collisions can also produce other long-lived hidden sector particles thus extending the possibilities for direct detection coverage of light sectors beyond dark matter.

The purpose of this work is to provide a first estimate of the dark matter flux from the aforementioned cosmic ray mechanism, taking into account its attenuation through the Earth. Using this estimate, we then place current and projected limits on the dark matter-nucleon cross-section from XENON1T and LZ data. We do this generally for a model-independent parametrisation of spin-independent cross-section vs dark matter mass and vs the meson branching ratio into dark matter. Finally, we consider a specific model in which the dark sector mediator is a hadrophilic scalar particle~\cite{Batell:2018fqo}. 


{\it Cosmic ray dark matter flux.} --- We distinguish two possible sources for a dark matter flux arising from the mechanism described earlier: inelastic cosmic ray collisions with protons in the interstellar medium and with the atmosphere on Earth. According to our calculations, the former yields a flux several orders of magnitude lower than the latter; therefore, we may safely neglect it and focus on our modelling of interactions at the atmosphere. 

The incoming cosmic ray flux is taken to be the local interstellar proton spectrum parametrised as in Ref.~\cite{Boschini:2017fxq}; alternatively we have also checked that using the AMS02 spectrum~\cite{Consolandi:2014uia} leads to identical results. The differential intensity $dI/dP$ as a function of particle rigidity $P$ is converted to a flux $d\Phi_p / dT_p = 2\pi (dP/dT_p)(dI/dP)$ per unit area, time, and kinetic energy $T_p$, over a hemispherical solid angle. We performed a Monte-Carlo simulation of this incoming flux using EPOS-LHC~\cite{Pierog:2013ria}, as implemented in the CRMC package~\cite{CRMC}, to simulate the collisions assuming the atmospheric nuclei target to be nitrogen at rest. The resulting $\pi^0$ and $\eta$ mesons~\footnote{We do not consider in this analysis charged mesons like $\pi^{\pm}$ or $K^{\pm}$ that are also abundantly produced in the atmosphere. The reason being the need of additional charged particles/leptonic decays that make our results model dependent and/or highly constrained branching ratios.} undergo two subsequent two-body decays via a vector or scalar mediator~\footnote{Here we consider only on-shell mediators though the sensitivity could in principle be extended to heavier off-shell mediators.} to a pair of dark matter particles, with a branching ratio that we keep as a free parameter. The rate of interactions is then integrated as follows over the volume of the atmosphere to obtain the total dark matter flux at the detector. 

The differential cosmic ray flux gets attenuated through the atmosphere as a function of height $h$ from ground level: 
\begin{equation}
\frac{d}{dh}\left(\frac{d\Phi_p}{dT_p}\right) = \sigma_{pN}(T_p) n_{\rm air}(h) \frac{d\Phi_p}{dT_p} \, ,
\label{eq:diff}
\end{equation}
where $\sigma_{pN}$ is the inelastic proton-nitrogen cross-section and $n_{\rm air}$ is the number density of air, taken from Ref.~\cite{Jursa:1985}, which is assumed to be entirely nitrogen for simplicity. Eq.~\ref{eq:diff} neglects higher order effects such as regenerations and secondary scatterings involved in a detailed cosmic ray shower model, but is sufficient to provide a conservative estimate of our hidden sector flux as including secondary meson production can only increase the flux. Since $\sigma_{pN} \simeq 255$ mb is constant to a good approximation over the relevant energy range, we may write $\frac{d\Phi_p}{dT_p}(T_p,h) \equiv y(h) \cdot \frac{d\Phi_p}{dT_p}(T_p)$ and solve for the attenuation factor $y(h)$. The dark matter flux at a detector located at a depth $z_d$ below ground is then given by
\begin{align}
\frac{d\Phi_\chi}{dT_\chi} &= \int_{R_E}^{R_E+h} R^2 dR \int_0^{2\pi}d\phi \int_{\cos\theta_\text{max}(T_\chi)}^1 \frac{d(cos{\theta})}{2\pi l(R,\theta, z_d)^2} \nonumber \\
& \times y(R^\prime) \int_{T_p^\text{min}}^{T_p^\text{max}} dT_p \frac{d\Phi_p}{dT_p}  n_{\rm air}(R^\prime) \frac{\sigma_{pN}}{\Gamma_{pN}} \frac{d\Gamma_{pN \to M \to \chi\chi}}{dT_p} \, , \nonumber \\
& \equiv \int_{T_p^\text{min}}^{T_p^\text{max}} dT_p \frac{d\Phi_p}{dT_p} n_{\rm air}^0 H_\text{eff}\frac{\sigma_{pN}}{\Gamma_{pN}} \frac{d\Gamma_{pN \to M \to \chi\chi}}{dT_p}
\label{eq:integratedflux}
\end{align}
where $R^\prime \equiv R-R_E$, $R_E$ is the radius of the Earth and $\theta_\text{max}(T_\chi)$ is a maximum angle dependent on the path length attenuation through the Earth, as described in the next Section. The line of sight distance $l(R,\theta, z_d)$ is given by 
\begin{equation}
l^2 = (R_E - z_d)^2 + R^2 - 2(R_E - z_d)R \cos{\theta} \, .
\end{equation}
It determines the rate dilution factor in the emission from source to detector that we have conservatively assumed to be isotropically distributed over a hemisphere. In the last line of \cref{eq:integratedflux} we defined an equivalent effective height at a constant number density taken to be the ground-level value, $n_{\rm air}^0 \simeq 5\times 10^{19} \text{ cm}^{-3}$~\cite{Jursa:1985}. We also define $H_\text{eff}$ as an effective height encoding the geometry of the problem. For example, with $\cos\theta_\text{max} = -1$, i.e. the Earth completely transparent to dark matter, we obtain $H_\text{eff} \simeq 5$ km. Finally $\Gamma_{pN \to M \to \chi\chi}/\Gamma_{pN}$ is a model-dependent branching ratio into the meson $M$ decaying to dark matter particles.

\begin{figure}
\begin{center}
\includegraphics[width=0.45 \textwidth]{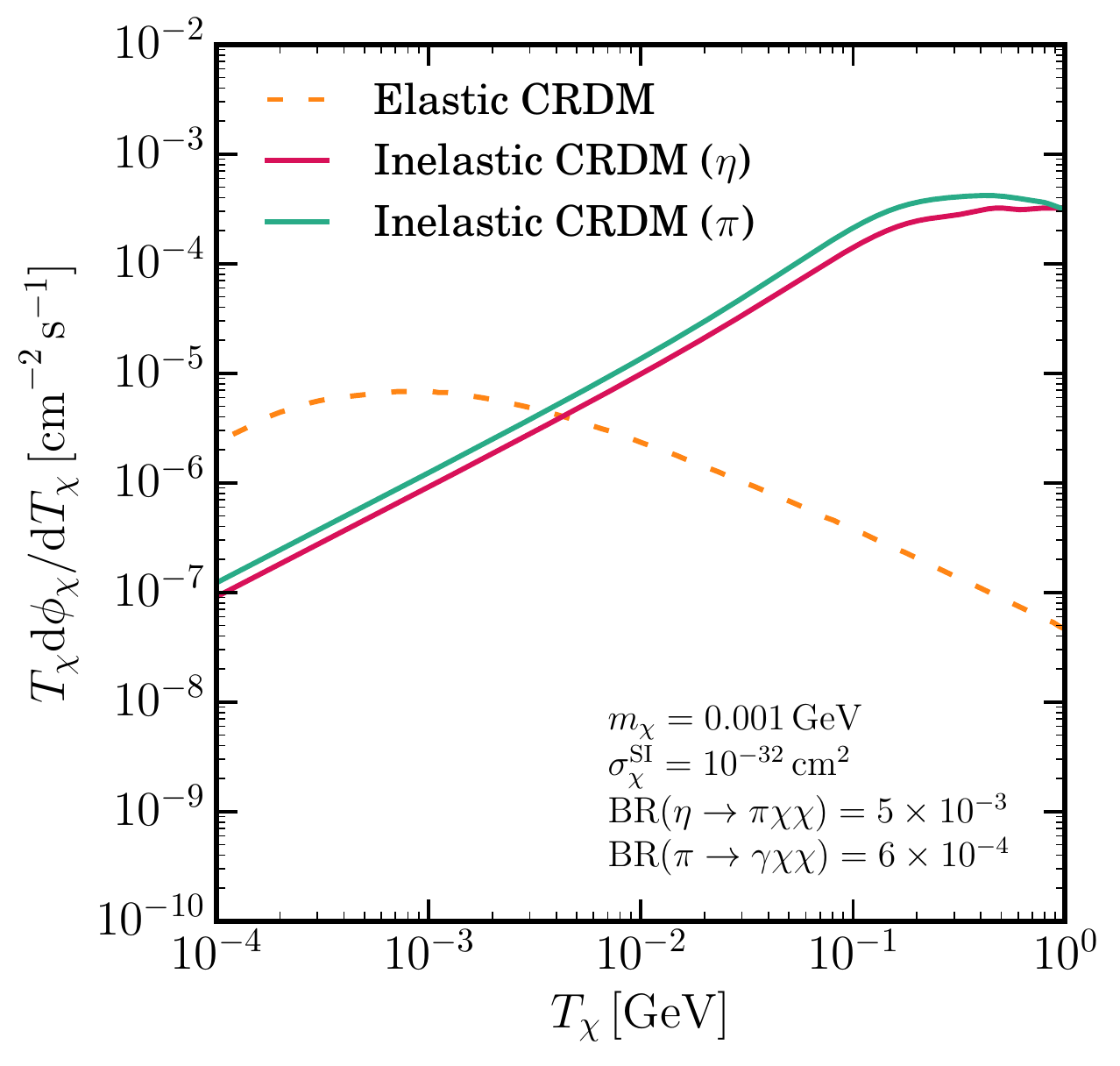}
\end{center}
\caption{Dark matter flux from cosmic rays for elastic collisions in dotted orange with $m_\chi = 1$ MeV, and for inelastic collisions with $BR(\pi^0 \to \gamma \chi \chi) = 6 \times 10^{-4}$ in solid green and $BR(\eta \to \pi\chi\chi) = 5 \times 10^{-3}$ in solid red.
\label{fig:flux}}
\end{figure}

The resulting dark matter flux in the transparent Earth case is plotted in \cref{fig:flux} in solid red for $BR(\eta \to \pi^0\chi\chi) = 4.6\times10^{-3}$ and green for $BR(\pi^0 \to \gamma\chi\chi) = 6\times 10^{-4}$, close to their experimental upper limits\footnote{There are currently no dedicated searches for $\eta \to \pi^0+\textrm{invisible}$. We assume here that the process $\eta \to \pi^0\chi\chi$ accounts for the uncertainties in the known $\eta$ decays.} \cite{Tanabashi:2018oca}. The fluxes are rather insensitive to the mediator and dark matter masses when these are produced on-shell. For comparison, we show in dotted orange the up-scattered flux for $m_\chi = 1$ MeV coming from elastic collisions of cosmic rays with interstellar dark matter, calculated as in Ref.~\cite{Bringmann:2018cvk}. Finally we have also checked that, when restricting to an opaque Earth, the muon flux in the 1-10 GeV range obtained in our approach agrees within an order of magnitude with data~\cite{Haino:2004nq,Tanabashi:2018oca}. 


{\it Attenuation through Earth.} --- As dark matter travels from the point of production through Earth a large enough nucleus interaction cross-section can prevent it from reaching the detector. The mean free path length together with the line of sight distance through Earth to the detector then determines the maximum polar angle at which we cut off the atmospheric volume integral in \cref{eq:integratedflux}. This line of sight distance through Earth is given by
\begin{align}
l_E &= \frac{1}{2}\left( b + \sqrt{b^2 + 4(R_E^2 - (R_E - z_d)^2)} \right) \, , \nonumber \\
b &\equiv \text{Sign}\left[R_E - z_d - (R_E + h)\cos\theta \right]  \nonumber \\
& \times 2(R_E - z_d) \sqrt{1 - \frac{(R_E +h)^2 \sin^2\theta}{l^2}} \, .
\end{align}

The mean free path length is determined by solving for the kinetic energy loss assuming a uniform distribution of nuclear recoil energy in elastic scattering, $d\sigma_{\chi N}/dT_r = \sigma_{\chi N} / T_r^\text{max}$, following Ref.~\cite{Bringmann:2018cvk}. Summing over the nuclei $N$, we then have
\begin{align}
\frac{dT_\chi}{dz} &= - \sum_N n_N \int_0^{T_r^\text{max}} \frac{d\sigma_{\chi N}}{dT_r} T_r dT_r \nonumber \\
&\underset{(T_\chi \ll m_N) }{\simeq} -\frac{1}{2m_\chi L}\left(T_\chi^2 + 2 m_\chi T_\chi \right) \, ,
\end{align}
where we used
\begin{equation}
T_r^\text{max} = \frac{T_\chi^2 + 2m_\chi T_\chi}{T_\chi + (m_\chi + m_N)^2/(2m_N)} \, ,
\end{equation}
and defined the mean free path length
\begin{equation}
L \equiv \left( \sum_N n_N \sigma_{\chi N} \frac{2 m_N m_\chi}{(m_\chi + m_N)^2} \right)^{-1} \, .
\end{equation}
Integrating this equation gives the incoming energy $T_\chi^0$ that is required to arrive at the detector with energy $T_\chi^z$ at distance $l_E$ through Earth:
\begin{equation}
T_\chi^0 = \frac{2 m_\chi T_\chi^z e^{l_E/L}}{2 m_\chi + T_\chi^z(1-e^{l_E/L})} \, .
\end{equation}
From this we obtain $\theta_\text{max}$ when $T_\chi^0 \to \infty$. The mean free path length is calculated by summing over the average number density of the elements given in Table 2 of Ref.~\cite{Kavanagh:2016pyr} which was originally presented in Ref.~\cite{Lundberg:2004dn}. We relate the nuclear interaction cross-section to the per nucleon spin-independent cross-section $\sigma_\chi^\text{SI}$ as
\begin{equation}
\sigma_{\chi N} = \sigma_\chi^\text{SI} A^2 \left(\frac{m_N}{m_p} \frac{(m_\chi + m_p)}{(m_\chi + m_N)} \right)^2 \, .
\end{equation} 

In practice we find that at the depth of the XENON1T detector the attenuation starts cutting off the atmospheric volume integral for cross-sections above $\sigma_\chi^\text{SI} \gtrsim 10^{-32} \text{ cm}^2$, with transmission falling exponentially above $\sim 10^{-28} \text{ cm}^2$. 


{\it Limits.} --- Finally, we obtain the expected rate at a detector coming from our inelastic cosmic ray dark matter flux by integrating within the detector nuclear recoil thresholds $T_1$ and $T_2$:
\begin{equation}
\Gamma_N = N_T \int_{T_1}^{T_2} dT_N \int_{T_\chi^\text{min}(T_N)}^\infty dT_\chi \epsilon(T_N) \frac{d\Phi_\chi}{dT_\chi} \frac{d\sigma_{\chi N}}{dT_N} \, ,
\end{equation}
where $N_T$ is the number of target atoms, $\epsilon$ is the detector nuclear recoil energy efficiency, and
\begin{equation}
T_\chi^\text{min} = \left(\frac{T_N}{2} - m_\chi \right)\left(1 \pm \sqrt{1 + \frac{2T_N}{m_N}\frac{(m_\chi + m_N)^2}{(2m_\chi - T_N)^2}} \right) \, ,
\end{equation}
with a plus sign if $T_\chi > 2m_N$ and minus sign otherwise. As an illustrative example we will focus on the limits from XENON1T. Its nuclear recoil energy threshold window is from $T_1 = 4.9$ keV to $T_2 = 40.9$ keV and the detector is located at a water-equivalent depth of 3.6 km, corresponding to 1.4 km of rock~\cite{Aprile:2018dbl}. For the 90\% CL limits we require a total number of events $N_{90\% \text{ CL}} = 3.56$ for the full exposure of 278.8 days of data collection with 1.3t fiducial mass. This event count, in Table 1 of Ref.~\cite{Aprile:2018dbl}, is the best fit given by a likelihood analysis for a 200 GeV WIMP whose recoil spectrum is comparable to that of the energetic light dark matter flux. 

\begin{figure}
\begin{center}
\includegraphics[width=0.49 \textwidth]{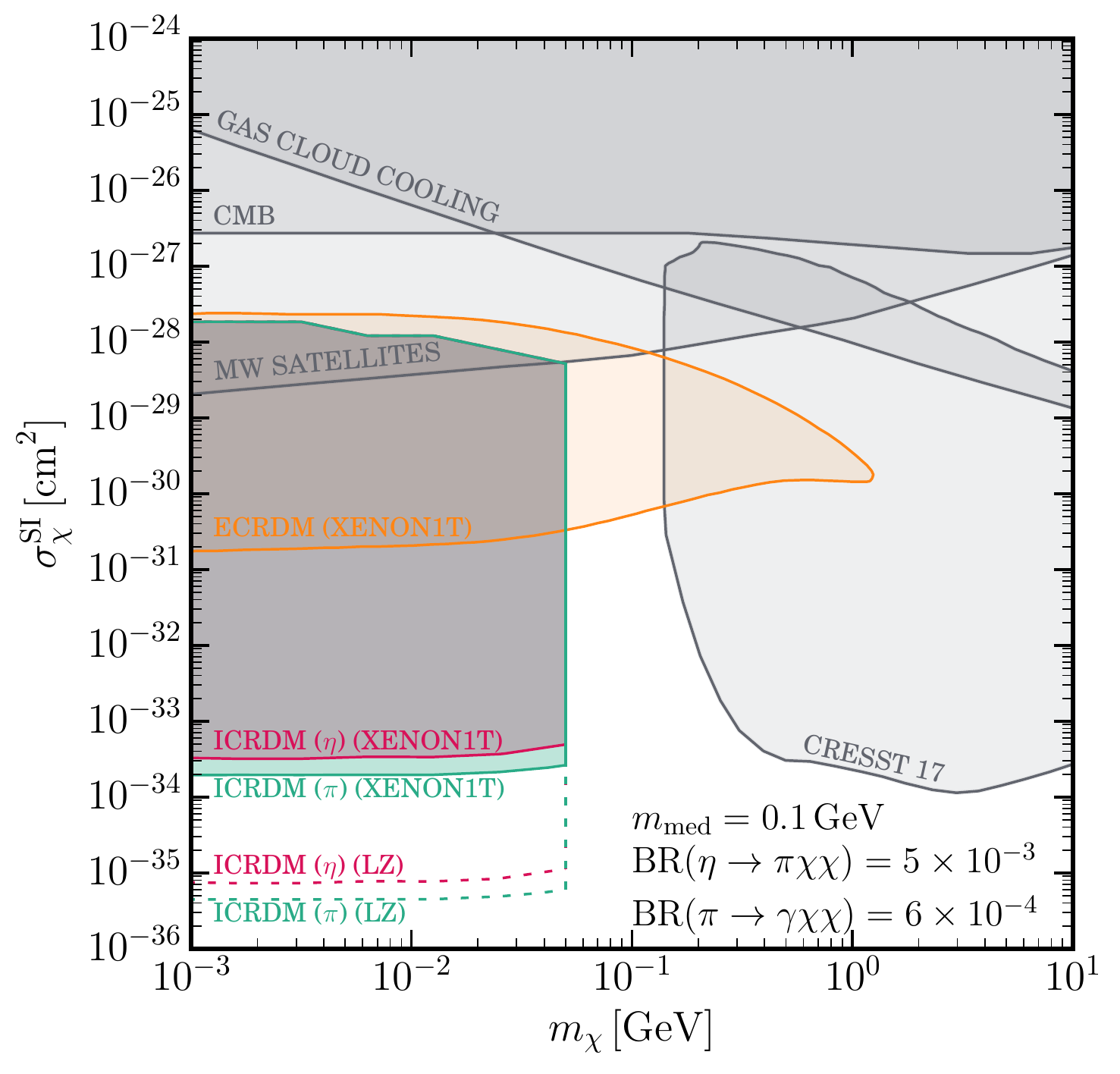} \\
\hspace{-0.35cm}\includegraphics[width=0.49 \textwidth]{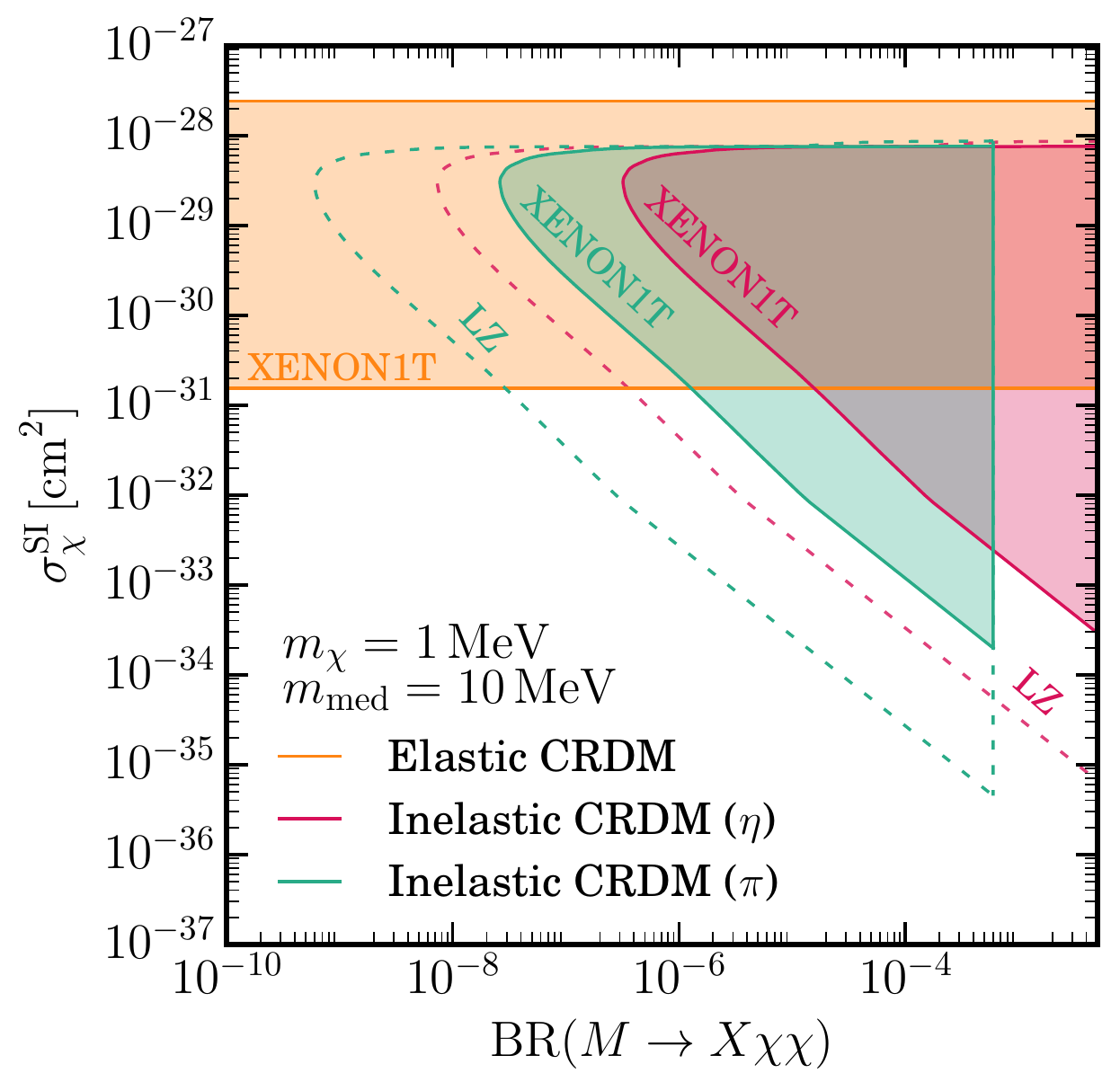}
\end{center}
\caption{90\% CL limits on the spin-independent dark matter-nucleon cross-section as a function of dark matter mass for a fixed branching ratio (\textit{top}) and as a function of branching ratio for a fixed dark matter mass (\textit{bottom}), as labelled. The inelastic cosmic ray dark matter limits from XENON1T~\cite{Aprile:2018dbl} are indicated in red and green for the flux originating from meson $M=\eta$ and $\pi^0$ decays, respectively, and in orange for elastic cosmic ray dark matter. The dashed lines are projections for the future LZ experiment~\cite{Akerib:2018lyp}. Other limits in grey are taken from Ref.~\cite{Bringmann:2018cvk} (based on CRESST~\cite{Angloher:2017sxg}, CMB~\cite{Xu:2018efh}, and gas cloud cooling~\cite{Bhoonah:2018wmw}), and from Milky-way satellites~\cite{Nadler:2019zrb}. 
\label{fig:limits}}
\end{figure}

For comparison with Ref.~\cite{Bringmann:2018cvk}, we first assume a uniform recoil energy distribution, $d\sigma_{\chi N}/dT_N = \sigma_{\chi N} / T_{r,N}^\text{max}$, and similarly obtain the resulting XENON1T limits on $\sigma_\chi^\text{SI}$. This is plotted in \cref{fig:limits} as a function of the dark matter mass for a fixed mediator mass and branching ratio values (\textit{top}), and as a function of the meson branching ratio into dark matter for a fixed mediator and dark matter mass (\textit{bottom}). The 90\% CL limits on inelastic cosmic ray dark matter from $\pi^0$ and $\eta$ decays are shown in green and red, respectively. As mentioned previously, the dark matter flux is relatively insensitive to their masses when these are light enough to be produced on-shell. We note that despite the rate of neutral pion production being an order of magnitude larger than for $\eta$ mesons, the relevant branching ratio of pions is experimentally constrained to be an order of magnitude smaller, as stated previously. The projected limits for the future LZ experiment~\cite{Akerib:2018lyp} are shown as dashed lines; we see that they improve on the cross-section sensitivity by almost two orders of magnitude. The corresponding limits from the irreducible flux of up-scattered dark matter for $m_\chi = 1$ MeV is given by the orange band and is independent of branching ratio. However, there is a (model-dependent) relation between the two---a dark matter coupling to nucleons will generically induce meson decay into dark matter, if kinematically allowed.

\begin{figure}
\begin{center}
\includegraphics[width=0.45 \textwidth]{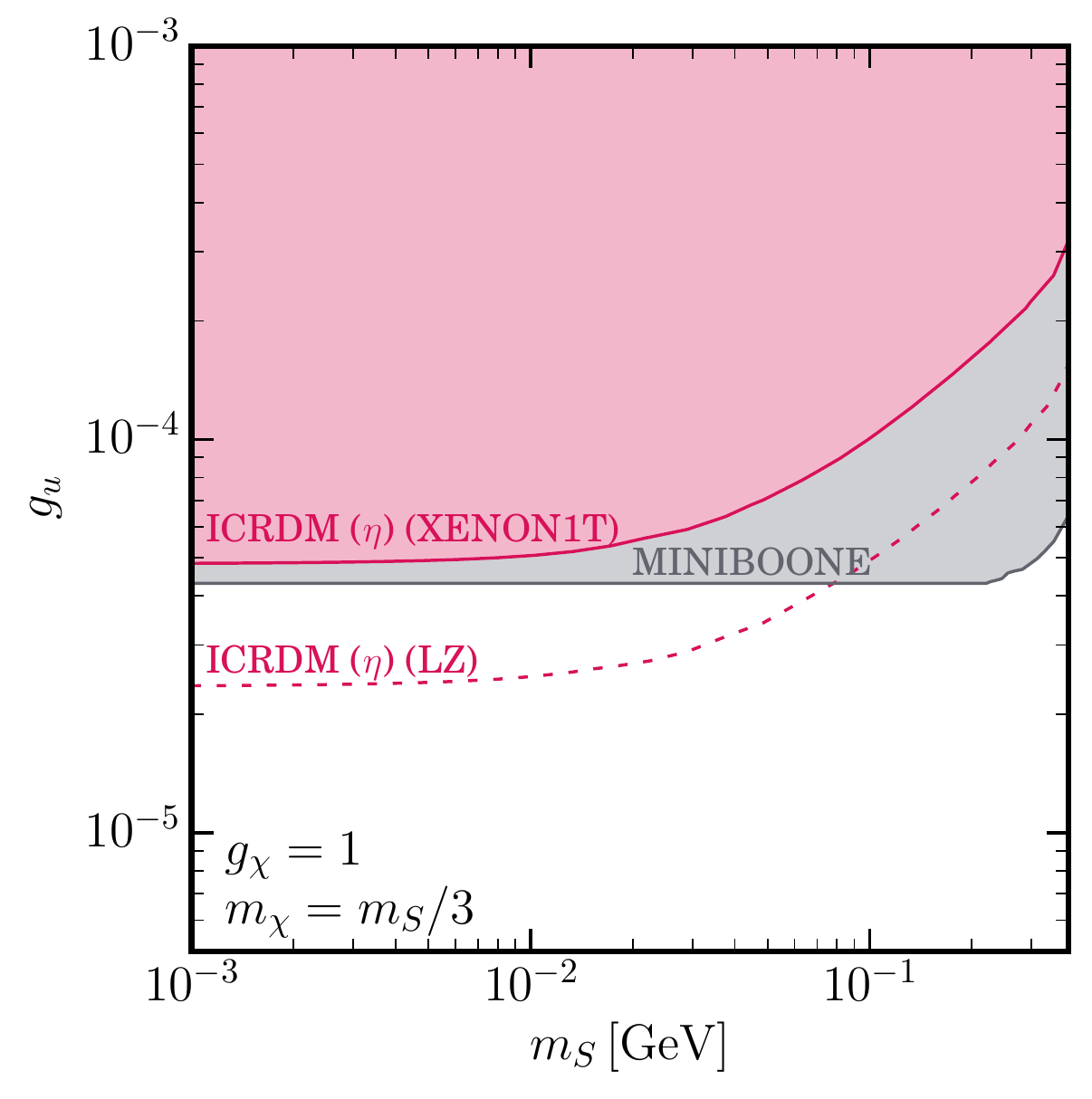}
\end{center}
\caption{90\% CL limits from inelastic cosmic ray dark matter flux from $\eta$ decays in red, for a hadrophilic scalar mediator of mass $m_S$ with up-quark coupling $g_u$, setting $g_\chi=1, m_\chi = m_S/3$. The solid line denotes current limits from XENON1T~\cite{Aprile:2018dbl}; the dashed line are future projections for the LZ experiment~\cite{Akerib:2018lyp}. Current MiniBoone limits from Ref.~\cite{Aguilar-Arevalo:2018wea} are shown in grey.
\label{fig:hadrophilic}}
\end{figure}

Next, we consider the hadrophilic scalar mediator model of Ref.~\cite{Batell:2018fqo}. The singlet scalar $S$ couples to a Dirac fermion dark matter $\chi$ and to the up quark through the Lagrangian terms
\begin{align}
\mathcal{L} \supset -g_\chi S\bar{\chi}_L \chi_R - g_u S \bar{u}_L u_R + \text{h.c.} \, .
\end{align}
The couplings to other flavours are assumed to be sub-dominant, so that we are left with four free parameters characterising the simplified model: $m_S$, $m_\chi$, $g_u$, and $g_\chi$. The branching ratio of $\eta$ mesons decaying into dark matter is given by 
\begin{equation}
BR(\eta \to \pi^0 S) = \frac{C^2 g_u^2 B^2}{16\pi m_\eta \Gamma_\eta} \lambda^{1/2}\left(1, \frac{m_S^2}{m_\eta^2}, \frac{m_\pi^2}{m_\eta^2}\right) \, ,
\end{equation}
where $B \simeq m_\pi^2/(m_u+m_d)$, $C \equiv \sqrt{1/3} \cos\theta^\prime -\sqrt{2/3} \sin\theta^\prime$ with $\theta^\prime \simeq -20^\circ$ and $\lambda(a,b,c) = a^2 + b^2 + c^2 -2ab - 2bc - 2ac$. We assume here that $BR(S \to \chi \chi) = 1$. For the differential $\chi$-nucleus cross-section~\footnote{Note that we assumed a uniform distribution of recoil energy in the Earth attenuation calculation, for analytical convenience, whereas we relax that assumption here. The Earth attenuation only becomes relevant for cross-sections above $\sim 10^{-28} \text{ cm}^2$ so this does not affect our limits.} involving a scalar mediator we have 
\begin{align}
\frac{d\sigma_{\chi N}}{dT_N} &= \frac{\left(Z y_{Spp} + (A-Z) y_{Snn}\right)^2 g_\chi^2}{8\pi} \nonumber \\
& \times \frac{(2m_N + T_N) (2 m_\chi^2 + m_N T_N)}{(T_\chi^2 + 2m_\chi T_\chi)(2m_N T_N + m_S^2)^2} F_H^2(\sqrt{2 m_N T_N}) \, ,
\end{align}
where $Z$ ($A-Z$) are the number of protons (neutrons), $y_{Spp} = 0.014 \cdot g_u m_p/m_u$, $y_{Snn} = 0.012 \cdot g_u m_n/m_u$, and $F_H$ is the Helm form factor~\cite{Duda:2006uk}. Computing the rate as described above, we obtain the 90\% CL limits shown in \cref{fig:hadrophilic} in red on the $g_u$ vs $m_S$ plane, for $g_\chi = 1, m_\chi = m_S/3$. The Earth suppresses the flux significantly only for values of $g_u$ greater than displayed. The MiniBoone limits from Ref.~\cite{Batell:2018fqo} are shown in grey. Note that for $g_\chi=1$ the constraints from the E787/E949 experiment are stronger than the MiniBoone and XENON1T limits~\cite{Batell:2018fqo}; however, they are set by invisible Kaon decays and are independent of $g_\chi$, whereas direct detection constraints from $\eta$ decay sources will grow quadratically with the dark matter coupling.  


{\it Conclusion.} --- As the search for dark matter broadens, it is becoming increasingly important to maximise every resource that we have, both technological and astrophysical. In this respect cosmic rays provide a valuable tool. It has long been appreciated that cosmic rays are a natural accelerator for probing high energies, while also being a source of background to indirect signals of dark matter decay; here we studied the potential of cosmic rays as a {\it source} of dark matter for direct detection. This opens up the potential of extending the sensitivity of various experiments to explore complementary parameter space, as we have illustrated for the case of XENON1T. It is remarkable that in this example the resulting limits are comparable to dedicated beam dump experiments such as MiniBoone. These limits will improve in the future with the LZ experiment, by about two orders of magnitude. It is also worth mentioning that the mechanism proposed in this work can generally be used as a source for any kind of light long-lived particle, not necessarily dark matter, although its detection will depend on its specific properties. In forthcoming work we plan to study such long-lived hidden sectors decaying back to Standard Model particles as well as the sensitivity of neutrino detectors.


\bigskip
{\it Acknowledgements.} --- We thank Christopher McCabe for valuable discussions on the XENON1T direct detection limits. TY is supported by a Branco Weiss Society in Science Fellowship and a Research Fellowship from Gonville and Caius College, and partially supported by STFC consolidated grant ST/P000681/1.  MF and MC are funded by the European  Research  Council  under  the  European Union's Horizon 2020 programme (ERC Grant Agreement no.648680 DARKHORIZONS).  In addition, the work of MF was supported partly by the STFC Grant ST/P000258/1.  JA is a recipient of an STFC quota studentship.

\bibliographystyle{apsrev4-1}
\bibliography{ICRDM}

\begin{thebibliography}{29}%
\makeatletter
\providecommand \@ifxundefined [1]{%
 \@ifx{#1\undefined}
}%
\providecommand \@ifnum [1]{%
 \ifnum #1\expandafter \@firstoftwo
 \else \expandafter \@secondoftwo
 \fi
}%
\providecommand \@ifx [1]{%
 \ifx #1\expandafter \@firstoftwo
 \else \expandafter \@secondoftwo
 \fi
}%
\providecommand \natexlab [1]{#1}%
\providecommand \enquote  [1]{``#1''}%
\providecommand \bibnamefont  [1]{#1}%
\providecommand \bibfnamefont [1]{#1}%
\providecommand \citenamefont [1]{#1}%
\providecommand \href@noop [0]{\@secondoftwo}%
\providecommand \href [0]{\begingroup \@sanitize@url \@href}%
\providecommand \@href[1]{\@@startlink{#1}\@@href}%
\providecommand \@@href[1]{\endgroup#1\@@endlink}%
\providecommand \@sanitize@url [0]{\catcode `\\12\catcode `\$12\catcode
  `\&12\catcode `\#12\catcode `\^12\catcode `\_12\catcode `\%12\relax}%
\providecommand \@@startlink[1]{}%
\providecommand \@@endlink[0]{}%
\providecommand \url  [0]{\begingroup\@sanitize@url \@url }%
\providecommand \@url [1]{\endgroup\@href {#1}{\urlprefix }}%
\providecommand \urlprefix  [0]{URL }%
\providecommand \Eprint [0]{\href }%
\providecommand \doibase [0]{http://dx.doi.org/}%
\providecommand \selectlanguage [0]{\@gobble}%
\providecommand \bibinfo  [0]{\@secondoftwo}%
\providecommand \bibfield  [0]{\@secondoftwo}%
\providecommand \translation [1]{[#1]}%
\providecommand \BibitemOpen [0]{}%
\providecommand \bibitemStop [0]{}%
\providecommand \bibitemNoStop [0]{.\EOS\space}%
\providecommand \EOS [0]{\spacefactor3000\relax}%
\providecommand \BibitemShut  [1]{\csname bibitem#1\endcsname}%
\let\auto@bib@innerbib\@empty
\bibitem [{\citenamefont {Goodman}\ and\ \citenamefont
  {Witten}(1985)}]{Goodman:1984dc}%
  \BibitemOpen
  \bibfield  {author} {\bibinfo {author} {\bibfnamefont {M.~W.}\ \bibnamefont
  {Goodman}}\ and\ \bibinfo {author} {\bibfnamefont {E.}~\bibnamefont
  {Witten}},\ }\href {\doibase 10.1103/PhysRevD.31.3059} {\bibfield  {journal}
  {\bibinfo  {journal} {Phys. Rev.}\ }\textbf {\bibinfo {volume} {D31}},\
  \bibinfo {pages} {3059} (\bibinfo {year} {1985})},\ \bibinfo {note}
  {[325(1984)]}\BibitemShut {NoStop}%
\bibitem [{\citenamefont {Battaglieri}\ \emph {et~al.}(2017)\citenamefont
  {Battaglieri} \emph {et~al.}}]{Battaglieri:2017aum}%
  \BibitemOpen
  \bibfield  {author} {\bibinfo {author} {\bibfnamefont {M.}~\bibnamefont
  {Battaglieri}} \emph {et~al.},\ }in\ \href
  {http://lss.fnal.gov/archive/2017/conf/fermilab-conf-17-282-ae-ppd-t.pdf}
  {\emph {\bibinfo {booktitle} {{U.S. Cosmic Visions: New Ideas in Dark
  Matter}}}}\ (\bibinfo {year} {2017})\ \Eprint
  {http://arxiv.org/abs/1707.04591} {arXiv:1707.04591 [hep-ph]} \BibitemShut
  {NoStop}%
\bibitem [{\citenamefont {Bringmann}\ and\ \citenamefont
  {Pospelov}(2019)}]{Bringmann:2018cvk}%
  \BibitemOpen
  \bibfield  {author} {\bibinfo {author} {\bibfnamefont {T.}~\bibnamefont
  {Bringmann}}\ and\ \bibinfo {author} {\bibfnamefont {M.}~\bibnamefont
  {Pospelov}},\ }\href {\doibase 10.1103/PhysRevLett.122.171801} {\bibfield
  {journal} {\bibinfo  {journal} {Phys. Rev. Lett.}\ }\textbf {\bibinfo
  {volume} {122}},\ \bibinfo {pages} {171801} (\bibinfo {year} {2019})},\
  \Eprint {http://arxiv.org/abs/1810.10543} {arXiv:1810.10543 [hep-ph]}
  \BibitemShut {NoStop}%
\bibitem [{\citenamefont {Ema}\ \emph {et~al.}(2018)\citenamefont {Ema},
  \citenamefont {Sala},\ and\ \citenamefont {Sato}}]{Ema:2018bih}%
  \BibitemOpen
  \bibfield  {author} {\bibinfo {author} {\bibfnamefont {Y.}~\bibnamefont
  {Ema}}, \bibinfo {author} {\bibfnamefont {F.}~\bibnamefont {Sala}}, \ and\
  \bibinfo {author} {\bibfnamefont {R.}~\bibnamefont {Sato}},\ }\href@noop {}
  {\  (\bibinfo {year} {2018})},\ \Eprint {http://arxiv.org/abs/1811.00520}
  {arXiv:1811.00520 [hep-ph]} \BibitemShut {NoStop}%
\bibitem [{\citenamefont {Aprile}\ \emph {et~al.}(2018)\citenamefont {Aprile}
  \emph {et~al.}}]{Aprile:2018dbl}%
  \BibitemOpen
  \bibfield  {author} {\bibinfo {author} {\bibfnamefont {E.}~\bibnamefont
  {Aprile}} \emph {et~al.} (\bibinfo {collaboration} {XENON}),\ }\href
  {\doibase 10.1103/PhysRevLett.121.111302} {\bibfield  {journal} {\bibinfo
  {journal} {Phys. Rev. Lett.}\ }\textbf {\bibinfo {volume} {121}},\ \bibinfo
  {pages} {111302} (\bibinfo {year} {2018})},\ \Eprint
  {http://arxiv.org/abs/1805.12562} {arXiv:1805.12562 [astro-ph.CO]}
  \BibitemShut {NoStop}%
\bibitem [{\citenamefont {Cappiello}\ \emph {et~al.}(2019)\citenamefont
  {Cappiello}, \citenamefont {Ng},\ and\ \citenamefont
  {Beacom}}]{Cappiello:2018hsu}%
  \BibitemOpen
  \bibfield  {author} {\bibinfo {author} {\bibfnamefont {C.~V.}\ \bibnamefont
  {Cappiello}}, \bibinfo {author} {\bibfnamefont {K.~C.~Y.}\ \bibnamefont
  {Ng}}, \ and\ \bibinfo {author} {\bibfnamefont {J.~F.}\ \bibnamefont
  {Beacom}},\ }\href {\doibase 10.1103/PhysRevD.99.063004} {\bibfield
  {journal} {\bibinfo  {journal} {Phys. Rev.}\ }\textbf {\bibinfo {volume}
  {D99}},\ \bibinfo {pages} {063004} (\bibinfo {year} {2019})},\ \Eprint
  {http://arxiv.org/abs/1810.07705} {arXiv:1810.07705 [hep-ph]} \BibitemShut
  {NoStop}%
\bibitem [{\citenamefont {Kouvaris}\ and\ \citenamefont
  {Pradler}(2017)}]{Kouvaris:2016afs}%
  \BibitemOpen
  \bibfield  {author} {\bibinfo {author} {\bibfnamefont {C.}~\bibnamefont
  {Kouvaris}}\ and\ \bibinfo {author} {\bibfnamefont {J.}~\bibnamefont
  {Pradler}},\ }\href {\doibase 10.1103/PhysRevLett.118.031803} {\bibfield
  {journal} {\bibinfo  {journal} {Phys. Rev. Lett.}\ }\textbf {\bibinfo
  {volume} {118}},\ \bibinfo {pages} {031803} (\bibinfo {year} {2017})},\
  \Eprint {http://arxiv.org/abs/1607.01789} {arXiv:1607.01789 [hep-ph]}
  \BibitemShut {NoStop}%
\bibitem [{\citenamefont {Ibe}\ \emph {et~al.}(2018)\citenamefont {Ibe},
  \citenamefont {Nakano}, \citenamefont {Shoji},\ and\ \citenamefont
  {Suzuki}}]{Ibe:2017yqa}%
  \BibitemOpen
  \bibfield  {author} {\bibinfo {author} {\bibfnamefont {M.}~\bibnamefont
  {Ibe}}, \bibinfo {author} {\bibfnamefont {W.}~\bibnamefont {Nakano}},
  \bibinfo {author} {\bibfnamefont {Y.}~\bibnamefont {Shoji}}, \ and\ \bibinfo
  {author} {\bibfnamefont {K.}~\bibnamefont {Suzuki}},\ }\href {\doibase
  10.1007/JHEP03(2018)194} {\bibfield  {journal} {\bibinfo  {journal} {JHEP}\
  }\textbf {\bibinfo {volume} {03}},\ \bibinfo {pages} {194} (\bibinfo {year}
  {2018})},\ \Eprint {http://arxiv.org/abs/1707.07258} {arXiv:1707.07258
  [hep-ph]} \BibitemShut {NoStop}%
\bibitem [{\citenamefont {Dolan}\ \emph {et~al.}(2018)\citenamefont {Dolan},
  \citenamefont {Kahlhoefer},\ and\ \citenamefont {McCabe}}]{Dolan:2017xbu}%
  \BibitemOpen
  \bibfield  {author} {\bibinfo {author} {\bibfnamefont {M.~J.}\ \bibnamefont
  {Dolan}}, \bibinfo {author} {\bibfnamefont {F.}~\bibnamefont {Kahlhoefer}}, \
  and\ \bibinfo {author} {\bibfnamefont {C.}~\bibnamefont {McCabe}},\ }\href
  {\doibase 10.1103/PhysRevLett.121.101801} {\bibfield  {journal} {\bibinfo
  {journal} {Phys. Rev. Lett.}\ }\textbf {\bibinfo {volume} {121}},\ \bibinfo
  {pages} {101801} (\bibinfo {year} {2018})},\ \Eprint
  {http://arxiv.org/abs/1711.09906} {arXiv:1711.09906 [hep-ph]} \BibitemShut
  {NoStop}%
\bibitem [{\citenamefont {Kouvaris}(2015)}]{Kouvaris:2015nsa}%
  \BibitemOpen
  \bibfield  {author} {\bibinfo {author} {\bibfnamefont {C.}~\bibnamefont
  {Kouvaris}},\ }\href {\doibase 10.1103/PhysRevD.92.075001} {\bibfield
  {journal} {\bibinfo  {journal} {Phys. Rev.}\ }\textbf {\bibinfo {volume}
  {D92}},\ \bibinfo {pages} {075001} (\bibinfo {year} {2015})},\ \Eprint
  {http://arxiv.org/abs/1506.04316} {arXiv:1506.04316 [hep-ph]} \BibitemShut
  {NoStop}%
\bibitem [{\citenamefont {An}\ \emph {et~al.}(2018)\citenamefont {An},
  \citenamefont {Pospelov}, \citenamefont {Pradler},\ and\ \citenamefont
  {Ritz}}]{An:2017ojc}%
  \BibitemOpen
  \bibfield  {author} {\bibinfo {author} {\bibfnamefont {H.}~\bibnamefont
  {An}}, \bibinfo {author} {\bibfnamefont {M.}~\bibnamefont {Pospelov}},
  \bibinfo {author} {\bibfnamefont {J.}~\bibnamefont {Pradler}}, \ and\
  \bibinfo {author} {\bibfnamefont {A.}~\bibnamefont {Ritz}},\ }\href {\doibase
  10.1103/PhysRevLett.120.141801, 10.1103/PhysRevLett.121.259903} {\bibfield
  {journal} {\bibinfo  {journal} {Phys. Rev. Lett.}\ }\textbf {\bibinfo
  {volume} {120}},\ \bibinfo {pages} {141801} (\bibinfo {year} {2018})},\
  \bibinfo {note} {[Erratum: Phys. Rev. Lett.121,no.25,259903(2018)]},\ \Eprint
  {http://arxiv.org/abs/1708.03642} {arXiv:1708.03642 [hep-ph]} \BibitemShut
  {NoStop}%
\bibitem [{\citenamefont {Emken}\ \emph {et~al.}(2018)\citenamefont {Emken},
  \citenamefont {Kouvaris},\ and\ \citenamefont {Nielsen}}]{Emken:2017hnp}%
  \BibitemOpen
  \bibfield  {author} {\bibinfo {author} {\bibfnamefont {T.}~\bibnamefont
  {Emken}}, \bibinfo {author} {\bibfnamefont {C.}~\bibnamefont {Kouvaris}}, \
  and\ \bibinfo {author} {\bibfnamefont {N.~G.}\ \bibnamefont {Nielsen}},\
  }\href {\doibase 10.1103/PhysRevD.97.063007} {\bibfield  {journal} {\bibinfo
  {journal} {Phys. Rev.}\ }\textbf {\bibinfo {volume} {D97}},\ \bibinfo {pages}
  {063007} (\bibinfo {year} {2018})},\ \Eprint
  {http://arxiv.org/abs/1709.06573} {arXiv:1709.06573 [hep-ph]} \BibitemShut
  {NoStop}%
\bibitem [{\citenamefont {Akerib}\ \emph {et~al.}(2018)\citenamefont {Akerib}
  \emph {et~al.}}]{Akerib:2018lyp}%
  \BibitemOpen
  \bibfield  {author} {\bibinfo {author} {\bibfnamefont {D.~S.}\ \bibnamefont
  {Akerib}} \emph {et~al.} (\bibinfo {collaboration} {LUX-ZEPLIN}),\
  }\href@noop {} {\  (\bibinfo {year} {2018})},\ \Eprint
  {http://arxiv.org/abs/1802.06039} {arXiv:1802.06039 [astro-ph.IM]}
  \BibitemShut {NoStop}%
\bibitem [{\citenamefont {Aguilar-Arevalo}\ \emph {et~al.}(2018)\citenamefont
  {Aguilar-Arevalo} \emph {et~al.}}]{Aguilar-Arevalo:2018wea}%
  \BibitemOpen
  \bibfield  {author} {\bibinfo {author} {\bibfnamefont {A.~A.}\ \bibnamefont
  {Aguilar-Arevalo}} \emph {et~al.} (\bibinfo {collaboration} {MiniBooNE DM}),\
  }\href {\doibase 10.1103/PhysRevD.98.112004} {\bibfield  {journal} {\bibinfo
  {journal} {Phys. Rev.}\ }\textbf {\bibinfo {volume} {D98}},\ \bibinfo {pages}
  {112004} (\bibinfo {year} {2018})},\ \Eprint
  {http://arxiv.org/abs/1807.06137} {arXiv:1807.06137 [hep-ex]} \BibitemShut
  {NoStop}%
\bibitem [{\citenamefont {Batell}\ \emph {et~al.}(2018)\citenamefont {Batell},
  \citenamefont {Freitas}, \citenamefont {Ismail},\ and\ \citenamefont
  {Mckeen}}]{Batell:2018fqo}%
  \BibitemOpen
  \bibfield  {author} {\bibinfo {author} {\bibfnamefont {B.}~\bibnamefont
  {Batell}}, \bibinfo {author} {\bibfnamefont {A.}~\bibnamefont {Freitas}},
  \bibinfo {author} {\bibfnamefont {A.}~\bibnamefont {Ismail}}, \ and\ \bibinfo
  {author} {\bibfnamefont {D.}~\bibnamefont {Mckeen}},\ }\href@noop {} {\
  (\bibinfo {year} {2018})},\ \Eprint {http://arxiv.org/abs/1812.05103}
  {arXiv:1812.05103 [hep-ph]} \BibitemShut {NoStop}%
\bibitem [{\citenamefont {Boschini}\ \emph {et~al.}(2017)\citenamefont
  {Boschini} \emph {et~al.}}]{Boschini:2017fxq}%
  \BibitemOpen
  \bibfield  {author} {\bibinfo {author} {\bibfnamefont {M.~J.}\ \bibnamefont
  {Boschini}} \emph {et~al.},\ }\href {\doibase 10.3847/1538-4357/aa6e4f}
  {\bibfield  {journal} {\bibinfo  {journal} {Astrophys. J.}\ }\textbf
  {\bibinfo {volume} {840}},\ \bibinfo {pages} {115} (\bibinfo {year}
  {2017})},\ \Eprint {http://arxiv.org/abs/1704.06337} {arXiv:1704.06337
  [astro-ph.HE]} \BibitemShut {NoStop}%
\bibitem [{\citenamefont {Consolandi}(2014)}]{Consolandi:2014uia}%
  \BibitemOpen
  \bibfield  {author} {\bibinfo {author} {\bibfnamefont {C.}~\bibnamefont
  {Consolandi}} (\bibinfo {collaboration} {AMS 02}),\ }in\ \href
  {http://www.slac.stanford.edu/econf/C131112/papers/1402.0467.pdf} {\emph
  {\bibinfo {booktitle} {{Proceedings, 33rd International Cosmic Ray Conference
  (ICRC2013): Rio de Janeiro, Brazil, July 2-9, 2013}}}}\ (\bibinfo {year}
  {2014})\ p.\ \bibinfo {pages} {1265},\ \Eprint
  {http://arxiv.org/abs/1402.0467} {arXiv:1402.0467 [astro-ph.HE]} \BibitemShut
  {NoStop}%
\bibitem [{\citenamefont {Pierog}\ \emph {et~al.}(2015)\citenamefont {Pierog},
  \citenamefont {Karpenko}, \citenamefont {Katzy}, \citenamefont {Yatsenko},\
  and\ \citenamefont {Werner}}]{Pierog:2013ria}%
  \BibitemOpen
  \bibfield  {author} {\bibinfo {author} {\bibfnamefont {T.}~\bibnamefont
  {Pierog}}, \bibinfo {author} {\bibfnamefont {I.}~\bibnamefont {Karpenko}},
  \bibinfo {author} {\bibfnamefont {J.~M.}\ \bibnamefont {Katzy}}, \bibinfo
  {author} {\bibfnamefont {E.}~\bibnamefont {Yatsenko}}, \ and\ \bibinfo
  {author} {\bibfnamefont {K.}~\bibnamefont {Werner}},\ }\href {\doibase
  10.1103/PhysRevC.92.034906} {\bibfield  {journal} {\bibinfo  {journal} {Phys.
  Rev.}\ }\textbf {\bibinfo {volume} {C92}},\ \bibinfo {pages} {034906}
  (\bibinfo {year} {2015})},\ \Eprint {http://arxiv.org/abs/1306.0121}
  {arXiv:1306.0121 [hep-ph]} \BibitemShut {NoStop}%
\bibitem [{\citenamefont {Baus}\ \emph {et~al.}()\citenamefont {Baus},
  \citenamefont {Pierog},\ and\ \citenamefont {Ulrich}}]{CRMC}%
  \BibitemOpen
  \bibfield  {author} {\bibinfo {author} {\bibfnamefont {C.}~\bibnamefont
  {Baus}}, \bibinfo {author} {\bibfnamefont {T.}~\bibnamefont {Pierog}}, \ and\
  \bibinfo {author} {\bibfnamefont {R.}~\bibnamefont {Ulrich}},\ }\href
  {https://web.ikp.kit.edu/rulrich/crmc.html} {\emph {\bibinfo {title} {Cosmic
  Ray Monte Carlo (CRMC)}}}\BibitemShut {NoStop}%
\bibitem [{\citenamefont {Jursa}(1985)}]{Jursa:1985}%
  \BibitemOpen
  \bibfield  {author} {\bibinfo {author} {\bibfnamefont {A.~S.}\ \bibnamefont
  {Jursa}},\ }\href@noop {} {\emph {\bibinfo {title} {Handbook of geophysics
  and the space environment}}},\ \bibinfo {edition} {4th}\ ed.\ (\bibinfo
  {publisher} {Hanscom Air Force Base, Mass. : Air Force Geophysics Laboratory,
  Air Force Systems Command, United States Air Force},\ \bibinfo {year}
  {1985})\BibitemShut {NoStop}%
\bibitem [{\citenamefont {Tanabashi}\ \emph {et~al.}(2018)\citenamefont
  {Tanabashi} \emph {et~al.}}]{Tanabashi:2018oca}%
  \BibitemOpen
  \bibfield  {author} {\bibinfo {author} {\bibfnamefont {M.}~\bibnamefont
  {Tanabashi}} \emph {et~al.} (\bibinfo {collaboration} {Particle Data
  Group}),\ }\href {\doibase 10.1103/PhysRevD.98.030001} {\bibfield  {journal}
  {\bibinfo  {journal} {Phys. Rev.}\ }\textbf {\bibinfo {volume} {D98}},\
  \bibinfo {pages} {030001} (\bibinfo {year} {2018})}\BibitemShut {NoStop}%
\bibitem [{\citenamefont {Haino}\ \emph {et~al.}(2004)\citenamefont {Haino}
  \emph {et~al.}}]{Haino:2004nq}%
  \BibitemOpen
  \bibfield  {author} {\bibinfo {author} {\bibfnamefont {S.}~\bibnamefont
  {Haino}} \emph {et~al.},\ }\href {\doibase 10.1016/j.physletb.2004.05.019}
  {\bibfield  {journal} {\bibinfo  {journal} {Phys. Lett.}\ }\textbf {\bibinfo
  {volume} {B594}},\ \bibinfo {pages} {35} (\bibinfo {year} {2004})},\ \Eprint
  {http://arxiv.org/abs/astro-ph/0403704} {arXiv:astro-ph/0403704 [astro-ph]}
  \BibitemShut {NoStop}%
\bibitem [{\citenamefont {Kavanagh}\ \emph {et~al.}(2017)\citenamefont
  {Kavanagh}, \citenamefont {Catena},\ and\ \citenamefont
  {Kouvaris}}]{Kavanagh:2016pyr}%
  \BibitemOpen
  \bibfield  {author} {\bibinfo {author} {\bibfnamefont {B.~J.}\ \bibnamefont
  {Kavanagh}}, \bibinfo {author} {\bibfnamefont {R.}~\bibnamefont {Catena}}, \
  and\ \bibinfo {author} {\bibfnamefont {C.}~\bibnamefont {Kouvaris}},\ }\href
  {\doibase 10.1088/1475-7516/2017/01/012} {\bibfield  {journal} {\bibinfo
  {journal} {JCAP}\ }\textbf {\bibinfo {volume} {1701}},\ \bibinfo {pages}
  {012} (\bibinfo {year} {2017})},\ \Eprint {http://arxiv.org/abs/1611.05453}
  {arXiv:1611.05453 [hep-ph]} \BibitemShut {NoStop}%
\bibitem [{\citenamefont {Lundberg}\ and\ \citenamefont
  {Edsjo}(2004)}]{Lundberg:2004dn}%
  \BibitemOpen
  \bibfield  {author} {\bibinfo {author} {\bibfnamefont {J.}~\bibnamefont
  {Lundberg}}\ and\ \bibinfo {author} {\bibfnamefont {J.}~\bibnamefont
  {Edsjo}},\ }\href {\doibase 10.1103/PhysRevD.69.123505} {\bibfield  {journal}
  {\bibinfo  {journal} {Phys. Rev.}\ }\textbf {\bibinfo {volume} {D69}},\
  \bibinfo {pages} {123505} (\bibinfo {year} {2004})},\ \Eprint
  {http://arxiv.org/abs/astro-ph/0401113} {arXiv:astro-ph/0401113 [astro-ph]}
  \BibitemShut {NoStop}%
\bibitem [{\citenamefont {Angloher}\ \emph {et~al.}(2017)\citenamefont
  {Angloher} \emph {et~al.}}]{Angloher:2017sxg}%
  \BibitemOpen
  \bibfield  {author} {\bibinfo {author} {\bibfnamefont {G.}~\bibnamefont
  {Angloher}} \emph {et~al.} (\bibinfo {collaboration} {CRESST}),\ }\href
  {\doibase 10.1140/epjc/s10052-017-5223-9} {\bibfield  {journal} {\bibinfo
  {journal} {Eur. Phys. J.}\ }\textbf {\bibinfo {volume} {C77}},\ \bibinfo
  {pages} {637} (\bibinfo {year} {2017})},\ \Eprint
  {http://arxiv.org/abs/1707.06749} {arXiv:1707.06749 [astro-ph.CO]}
  \BibitemShut {NoStop}%
\bibitem [{\citenamefont {Xu}\ \emph {et~al.}(2018)\citenamefont {Xu},
  \citenamefont {Dvorkin},\ and\ \citenamefont {Chael}}]{Xu:2018efh}%
  \BibitemOpen
  \bibfield  {author} {\bibinfo {author} {\bibfnamefont {W.~L.}\ \bibnamefont
  {Xu}}, \bibinfo {author} {\bibfnamefont {C.}~\bibnamefont {Dvorkin}}, \ and\
  \bibinfo {author} {\bibfnamefont {A.}~\bibnamefont {Chael}},\ }\href
  {\doibase 10.1103/PhysRevD.97.103530} {\bibfield  {journal} {\bibinfo
  {journal} {Phys. Rev.}\ }\textbf {\bibinfo {volume} {D97}},\ \bibinfo {pages}
  {103530} (\bibinfo {year} {2018})},\ \Eprint
  {http://arxiv.org/abs/1802.06788} {arXiv:1802.06788 [astro-ph.CO]}
  \BibitemShut {NoStop}%
\bibitem [{\citenamefont {Bhoonah}\ \emph {et~al.}(2018)\citenamefont
  {Bhoonah}, \citenamefont {Bramante}, \citenamefont {Elahi},\ and\
  \citenamefont {Schon}}]{Bhoonah:2018wmw}%
  \BibitemOpen
  \bibfield  {author} {\bibinfo {author} {\bibfnamefont {A.}~\bibnamefont
  {Bhoonah}}, \bibinfo {author} {\bibfnamefont {J.}~\bibnamefont {Bramante}},
  \bibinfo {author} {\bibfnamefont {F.}~\bibnamefont {Elahi}}, \ and\ \bibinfo
  {author} {\bibfnamefont {S.}~\bibnamefont {Schon}},\ }\href {\doibase
  10.1103/PhysRevLett.121.131101} {\bibfield  {journal} {\bibinfo  {journal}
  {Phys. Rev. Lett.}\ }\textbf {\bibinfo {volume} {121}},\ \bibinfo {pages}
  {131101} (\bibinfo {year} {2018})},\ \Eprint
  {http://arxiv.org/abs/1806.06857} {arXiv:1806.06857 [hep-ph]} \BibitemShut
  {NoStop}%
\bibitem [{\citenamefont {Nadler}\ \emph {et~al.}(2019)\citenamefont {Nadler},
  \citenamefont {Gluscevic}, \citenamefont {Boddy},\ and\ \citenamefont
  {Wechsler}}]{Nadler:2019zrb}%
  \BibitemOpen
  \bibfield  {author} {\bibinfo {author} {\bibfnamefont {E.~O.}\ \bibnamefont
  {Nadler}}, \bibinfo {author} {\bibfnamefont {V.}~\bibnamefont {Gluscevic}},
  \bibinfo {author} {\bibfnamefont {K.~K.}\ \bibnamefont {Boddy}}, \ and\
  \bibinfo {author} {\bibfnamefont {R.~H.}\ \bibnamefont {Wechsler}},\
  }\href@noop {} {\  (\bibinfo {year} {2019})},\ \Eprint
  {http://arxiv.org/abs/1904.10000} {arXiv:1904.10000 [astro-ph.CO]}
  \BibitemShut {NoStop}%
\bibitem [{\citenamefont {Duda}\ \emph {et~al.}(2007)\citenamefont {Duda},
  \citenamefont {Kemper},\ and\ \citenamefont {Gondolo}}]{Duda:2006uk}%
  \BibitemOpen
  \bibfield  {author} {\bibinfo {author} {\bibfnamefont {G.}~\bibnamefont
  {Duda}}, \bibinfo {author} {\bibfnamefont {A.}~\bibnamefont {Kemper}}, \ and\
  \bibinfo {author} {\bibfnamefont {P.}~\bibnamefont {Gondolo}},\ }\href
  {\doibase 10.1088/1475-7516/2007/04/012} {\bibfield  {journal} {\bibinfo
  {journal} {JCAP}\ }\textbf {\bibinfo {volume} {0704}},\ \bibinfo {pages}
  {012} (\bibinfo {year} {2007})},\ \Eprint
  {http://arxiv.org/abs/hep-ph/0608035} {arXiv:hep-ph/0608035 [hep-ph]}
  \BibitemShut {NoStop}%
\end{thebibliography}%

\end{document}